\documentclass[10pt,letterpaper]{article}

\usepackage{graphicx}

\usepackage{color}
\usepackage{authblk}

\def\be{\begin{eqnarray}}
\def\ee{\end{eqnarray}}
\def\nn{\nonumber}

\def\anda{~{\rm and}~}

\title{Using graph theory to compute Laplace operators arising in a
  model  for blood flow in capillary networks}
\author[1]{D. Terman}
\author[2]{Y. Hannawi}
\affil[1]{Deparment of Mathematics, Ohio State University, Columbus, Ohio, USA}
\affil[2]{Division of Cerebrovascular Diseases and Neurocritical Care, Department of Neurology, The Ohio State University Wexner Medical Center, Columbus, Ohio, USA}
\date{\today}

\begin{document}
\maketitle

\begin{abstract}
 Maintaining cerebral blood flow is critical for
adequate neuronal function.
Previous computational models of
brain capillary networks have predicted that heterogeneous cerebral
capillary flow patterns result in lower brain tissue partial oxygen
pressures. It has been suggested that this may lead to number of
diseases such as Alzheimer’s disease, acute ischemic stroke, traumatic brain
injury and ischemic heart disease.
We have previously developed a computational model that was used to describe
in detail the effect of flow heterogeneities on tissue oxygen levels.
The main result in that paper  was
that, for a general class of capillary networks, perturbations of
 segment diameters or conductances always lead to decreased oxygen
 levels.  This result was varified using both numerical simulations and
 mathematical analysis. However, the analysis depended on a novel
 conjecture concerning the Laplace operator of functions related to
 the segment flow rates and how they depend on the conductances.
 The goal of this paper is to give a mathematically rigorous proof of
 the conjecture for a general class of  networks.
The proof depends on determining the number of trees and forests
in certain graphs arising from the capillary network.

  \end{abstract}

\section{Introduction}

Maintaining cerebral blood flow is critical for
adequate neuronal function \cite{Girouard2006,Iadecola2004}.
Previous computational models of
brain capillary networks have predicted that heterogeneous cerebral
capillary flow patterns result in lower brain tissue partial oxygen
pressures \cite{Jespersen2012}. It has been suggested that this may lead to number of
diseases
such as Alzheimer’s disease, acute ischemic stroke, traumatic brain
injury and ischemic heart disease \cite{Ostergaard2014a, Ostergaard2014b}.

In \cite{chen2020}, we developed a computational model that was used to describe
in detail the effect of flow heterogeneities on tissue oxygen levels.
The primary question addressed in \cite{chen2020} was: How do the oxygen levels
depend  on changes in network parameters such as segment diameters and conductances? In particular, if we
randomly perturb a given choice of parameters, will the oxygen
levels, on average, increase or decrease? The main result in \cite{chen2020}  was
that, for a general class of capillary networks, perturbations of
 segment diameters or conductances always lead to decreased oxygen
 levels.

 This result was varified using both numerical simulations and
 mathematical analysis. However, the analysis depended on a novel
 conjecture concerning the Laplace operator of functions related to
 the segment flow rates and how they depend on the conductances.
 The goal of this paper is to give a mathematically rigorous proof of
 the conjecture for a class of  networks.

 An outline of the paper is the following. In the next section, we
 state our  conjecture, as well as the main results. In Section 3,
 we  present the model for capillary blool flow
 developed in \cite{chen2020}  and discuss how results presented in this paper
 are related to those given in \cite{chen2020}. The results depend on computing the
 Laplace
 operator of certain functions, which depend on the flow rates. In
 Section 4, we compute explicit formulas for these Laplacians.
 In Section 5, we describe numerical simulations, which demonstrate
 that the conjecture holds for a general class of capillary networks.
 Finally, in Section 6, we rigorously prove that the conjecture holds
 for
 a specific class of networks. The proof depends on determining the number
of trees and forests in certain graphs arising from the capillary network.

\section{Statement of main results}
 
 Blood flow in brain capillary networks is often modeled using
an undirected, weighted graph.
Suppose that this graph has K nodes, which we denote as simply 1, 2,
  …, K. Each node has degree greater than one, except for nodes
corresponding
to where blood either enters or leaves the network. These nodes have
degree one. 

To each edge $e_{ij}$, connecting nodes $i$ and $j$,
we assign a conductance, $\alpha_{ij}$. Moreover,
to each node $i$, there corresponds a blood pressure, $P_i$. We
assume that the blood pressures at the incoming and outgoing nodes
are given. 
Then the remaining blood pressures are determined by
assuming conservation of blood
flow at each node. That is, the blood flow rate along some edge $e_{ij}$
 is given by
\be \label{flowrates}
Q_{ij} &=& \alpha_{ij} (P_i - P_j). 
\ee
We assume that for each node $i$, the sum of all the blood flow rates
along edges
from node $i$ is zero. This leads to a linear algebra problem
(which is described in detail later) for the remaining blood pressures.

For each edge $e_{ij}$,  let $\Gamma_{ij}= 1/Q_{ij}$. Note that  each
$\Gamma_{ij}$ is a function of all the conductances $ \alpha_{rs} $. Let
$\Delta$
be the Laplace
operator. That is,
\be
 \Delta \Gamma_{ij} &=& \sum_{r,s} \frac{\partial^2
   \Gamma_{ij}}{\partial \alpha_{rs}^2}. \nn
 \ee
 Our conjecture is then
 \medskip
 
 \noindent {\bf Main Result:} Let $e_{ij}$ be any edge with 
 $P_i -P_j>0$. Then $\Delta \Gamma_{ij}$, evaluated when all
 the conductances are equal, is positive. 
 \medskip

This result is varified for a general class of networks using
numerical simulations and for a specific class of networks using
rigorous mathematical analysis.

\section{Motivation of the Main Result}

Here we briefly describe the model for capillary blool flow
 developed in \cite{chen2020}  and discuss how results presented in this paper
 are related to those given in \cite{chen2020}.
 
We begin with a graph as described above, except we now assume that
there
is just one incoming node, at $N_{in}$.
We assign a conductance $\alpha_{ij}$
to each edge $e_{ij}$ and 
blood pressures to the incoming and outgoing nodes. We then
compute blood pressures, $P_i$, at all of the nodes and flow rates, $Q_{ij}$, along
each of the edges, as described above.

To each node, $i$, there also corresponds an oxygen partial pressure,
$\Omega_i$. These are determined as follows. We assume that
$\Omega_{Nin}$ is given
at the
incoming node.
Suppose that $e_{ij}$ is some edge with $P_i > P_j$
so that $Q_{ij} > 0$. We parameterize this edge by the distance, $x$,
from node $i$ and assume that along this edge,
$\Omega_{ij}(x)$ decays according to an equation of the form
\be \label{kmet}
\frac{d \Omega_{ij}}{dx} &=& -\frac{\rho}{Q_{ij}} \, F(\Omega_{ij}) 
\ee
with $\Omega_{ij}(0) = \Omega_i$. Here $\rho$ is a
fixed parameter  and $F$ is simply assumed to be a positive, smooth
function. We need some rule to determine how $\Omega_j$ is computed at
each node, $j$.
If there is just one node $i$
with $Q_{ij} > 0$, then $\Omega_j = \Omega_{ij}(l_{ij})$ where
$l_{ij}$ is the length of edge $e_{ij}$.
If there are two nodes $i$ and $k$ so that $Q_{ij} >0 \anda Q_{kj} >
0$, then
$\Omega_j \, = \, C_1 \, \Omega_{ij}(l_{ij}) + C_2 \, \Omega_{kj}(l_{kj})$
for some positive constants $C_1 \anda C_2$.

In \cite{chen2020}, we show that random perturbations of conductances
lead, on average, to a decrease in oxygen levels.
More precisely, suppose that the conductances along the  edge $e_{ij}$
are given by $\alpha_{ij}^0$. For $\epsilon > 0$, we say that $\{
\alpha_{ij} \}$ is an $\epsilon$-perturbation of $\{ \alpha_{ij}^0 \}$ if
$| \alpha_{ij} - \alpha_{ij}^0 | < \epsilon$ for each edge. We
assume that $0 < \epsilon < {\rm min} \{\alpha_{ij}^0 \}$. For a given
set of conducctances,  $\{\alpha_{ij} \}$, we can compute the oxygen
partial pressure $\Omega_i$ at each of the nodes. Let $\Omega_{i
  \epsilon}$ equal to the average oxygen partial pressure at node $i$
taken over all $\epsilon$-perturbations of $\{ \alpha_{ij}^0 \}$.

The
main result in \cite{chen2020}  is that if the  $\{ \alpha_{ij}^0 \}$ are all some
fixed constant and the parameter $\rho$,
which appears in (\ref{kmet}), is sufficiently small,
then
$\Omega_{i  \epsilon} < \Omega_{i0}$ at each node $i$.

This result was demonstrated  by noting that the oxygen
levels $\Omega_i$ are all functions of the conductances $\alpha_{rs}$.
We showed numerically in \cite{chen2020}  that for a general class of networks,
$\Delta \, \Omega_i \, < \, 0$ at each node. Here, $\Delta$ is the
Laplace
operater with respect to the conductance variables. It then follows
from the so-called Maximum Principle for the Laplace operator that
each $\Omega_i$ is greater than the average value of the oxygen levels
over all perturbations of the conductances of a fixed size; that is,
$\Omega_{i  \epsilon} < \Omega_{i0}$ at each node $i$.

A key step in the analysis of this result was to consider
$\Gamma_{ij} \, = \, 1/Q_{ij}$, as defined above. In \cite{chen2020}, we proved the
following
\medskip

\noindent {\bf Proposition}: If $\Delta \, \Gamma_{ij} \, > \, 0$ 
for each edge $e_{ij}$ with $P_i \, - \,  P_j \, > \, 0$ and $\rho$  is sufficiently small, then $\Delta \Omega_i \,
< \, 0$ for each node.
\medskip

Hence, the Main Result of this paper plays a central role in analyzing
the model presented in \cite{chen2020}. 
 
 \section{Computation of $\Delta \Gamma_{ij}$}

Here we assume that there is just one incoming node and one outgoing
node;
these are $N_{in}=K \anda N_{out}=K-1$, respectively. It is
straightforward
to extend the formulas which follow if there are multiple incoming or
outgoing nodes.
 
 Let $\cal{A}$ be the $(K-2) \times (K-2)$ matrix defined by
 \[ {\cal A}_{ij} = \left\{ \begin{array}{cl}
         -\alpha_{ij} &  {\rm if} ~~ i \neq j \\
         \sum_{k=1}^K \alpha_{ik} & {\rm if} ~~ i = j. \end{array}
\right. \]
Here, $\alpha_{ij} = 0$ if there is no edge connecting nodes $i \anda j$.
   Let
$I_{in} \anda I_{out}$ be the 
nodes that share edges with the incoming and outgoing nodes, respectively,  $P_{in} \anda P_{out}$ be the blood
pressures
at the incoming and outgoing nodes  and
$\cal{B}$ be the $(K-2) \times 1$ column matrix
\[ {\cal B}_{i,1} = \left\{ \begin{array}{cl}
                              
                              \alpha_{lin, N-1} ~P_{in} & {\rm if}~~ i
                                                         = I_{in} \\
                              \alpha_{lout, N} ~~P_{out} & {\rm if} ~~ i
                                                         = I_{out} \\
                              0 &  {\rm otherwise.} \end{array}
                          \right. \]
Then the blood pressures ${\cal P} \, = \,(P_1, P_2, ..., P_{K-2})$
 satisfy
${\cal A} \, {\cal P} \, = \, {\cal B}.$

We solve for the $P_i$ using Cramer's rule.
For each  $i$, with $1 \leq i \leq K-2$, let ${\cal D}^i$ be the
matrix in which the $i^{\rm th}$ column of $\cal A$ is replaced with
$\cal B$,  $\delta = {\rm det}~\cal A$ and  $\delta_i = {\rm
  det}~{\cal D}^i$.
Then $P_i ~=~  \delta_i /\delta. $

Note that $\delta$ and each $\delta_i$ are  linear functions of
the conductances, $ \alpha_{rs}$. Hence, for each conductance $\alpha_{rs}$, we
can write
\be
\delta  &=& \alpha_{rs} A^1_{rs} + A^0_{rs}  ~~ \anda ~~
\delta_i ~=~ \alpha_{rs} D^{i1}_{rs} + D^{i0}_{rs} \nn
\ee
where $ A^1_{rs}, \,  A^0_{rs}, \, D^{i1}_{rs} \anda D^{i0}_{rs}$ do not depend on $\alpha_{rs}$.
It follows that if $1 \leq i, j \leq K-2$, then for each
$\alpha_{rs}$,

\be
\Gamma_{ij} &=& \frac{1}{Q_{ij}} ~=~ \frac{1}{\alpha_{ij} (P_i - P_j)}
\nn \\ 
&=& \frac{1}{\alpha_{ij}} 
\left( \frac{ \alpha_{rs} A^1_{rs} + A^0_{rs} } 
{ \alpha_{rs}(D^{i1}_{rs} - D^{j1}_{rs}) + (D^{i0}_{rs}-D^{j0}_{rs})}
\right). \nn
\ee
If $\alpha_{ij} \neq \alpha_{rs}$, then
\be \label{dgamma1}
\frac{\partial^2 \, \Gamma_{ij}}{\partial \, \alpha^2_{rs}} &=&
-2 (D^{i1}_{rs} - D^{j1}_{rs}) \left( \frac{
    A^1_{rs}~(D^{i0}_{rs}-D^{j0}_{rs}) - A^0_{rs}~(D^{i1}_{rs} -
    D^{j1}_{rs}) }
{ \alpha_{ij} (\delta_i - \delta_j)^3} \right). \\ \nn
\ee
If $\alpha_{ij} =  \alpha_{rs}$, then $D_{rs}^{j1} = D_{rs}^{j1}$ and
\be \label{dgamma2}
\frac{\partial^2 \, \Gamma_{ij}}{\partial \, \alpha^2_{rs}} &=&
\frac{2 A^0_{rs}} {\alpha_{rs}^3 (\delta_i - \delta_j)}. \\ \nn
\ee
Now suppose that $i=K-1 \anda j=I_{in}$. Then $P_i = P_{in}$. 
If $\alpha_{ij} \neq \alpha_{rs}$, then
\be \label{dgamma3}
\frac{\partial^2 \, \Gamma_{ij}}{\partial \, \alpha^2_{rs}} &=&
-2 (P_{in} A^1_{rs} - D^{j1}_{rs}) \left(
\frac{A^1_{rs}~(P_{in} A^0_{rs}-D^{j0}_{rs}) - A^0_{rs}~(P_{in}A^{1}_{rs} - D^{j1}_{rs})}
{ \alpha_{ij}~( P_{in}  \delta - \delta_j)^3} \right). \nn \\
\ee
If $\alpha_{ij}=\alpha_{rs}$, then $P_{in}A^1_{rs}=D^{j1}_{rs}$ and
\be \label{dgamma4}
\frac{\partial^2 \Gamma_{ij}}{\partial \alpha^2_{rs}} &=& 
\frac{2 A^0_{rs}}{\alpha_{rs}^3 (P_{in} \delta -\delta_j)}. \\ \nn 
\ee
Finally, suppose that $i=I_{out} \anda j=K$. Then $P_j = P_{out}$.
If $\alpha_{ij} \neq \alpha_{rs}$, then
\be \label{dgamma5}
\frac{\partial^2 \, \Gamma_{ij}}{\partial \, \alpha^2_{rs}} &=&
-2(  D^{i1}_{rs}-P_{out} A^1_{rs}) \left(
\frac{A^1_{rs}~(D^{i0}_{rs}-P_{out} A^0_{rs}) - A^0_{rs}~( D^{i1}_{rs}-P_{out}A^{1}_{rs})}
{ \alpha_{ij}~( \delta_i- P_{out} \delta)^3} \right) \nn \\ 
\ee
If $\alpha_{ij}=\alpha_{rs}$, then $P_{out}A^1_{rs}=D^{j1}_{rs}$ and
\be \label{dgamma6}
\frac{\partial^2 \Gamma_{ij}}{\partial \alpha^2_{rs}} &=&
\frac{2 A^0_{rs}}{\alpha_{rs}^3 (\delta_i -P_{out} \delta)}.
\ee

\section{Voronoi Networks}

\begin{figure}
\centering
\includegraphics[width=3in,height=2.5in]{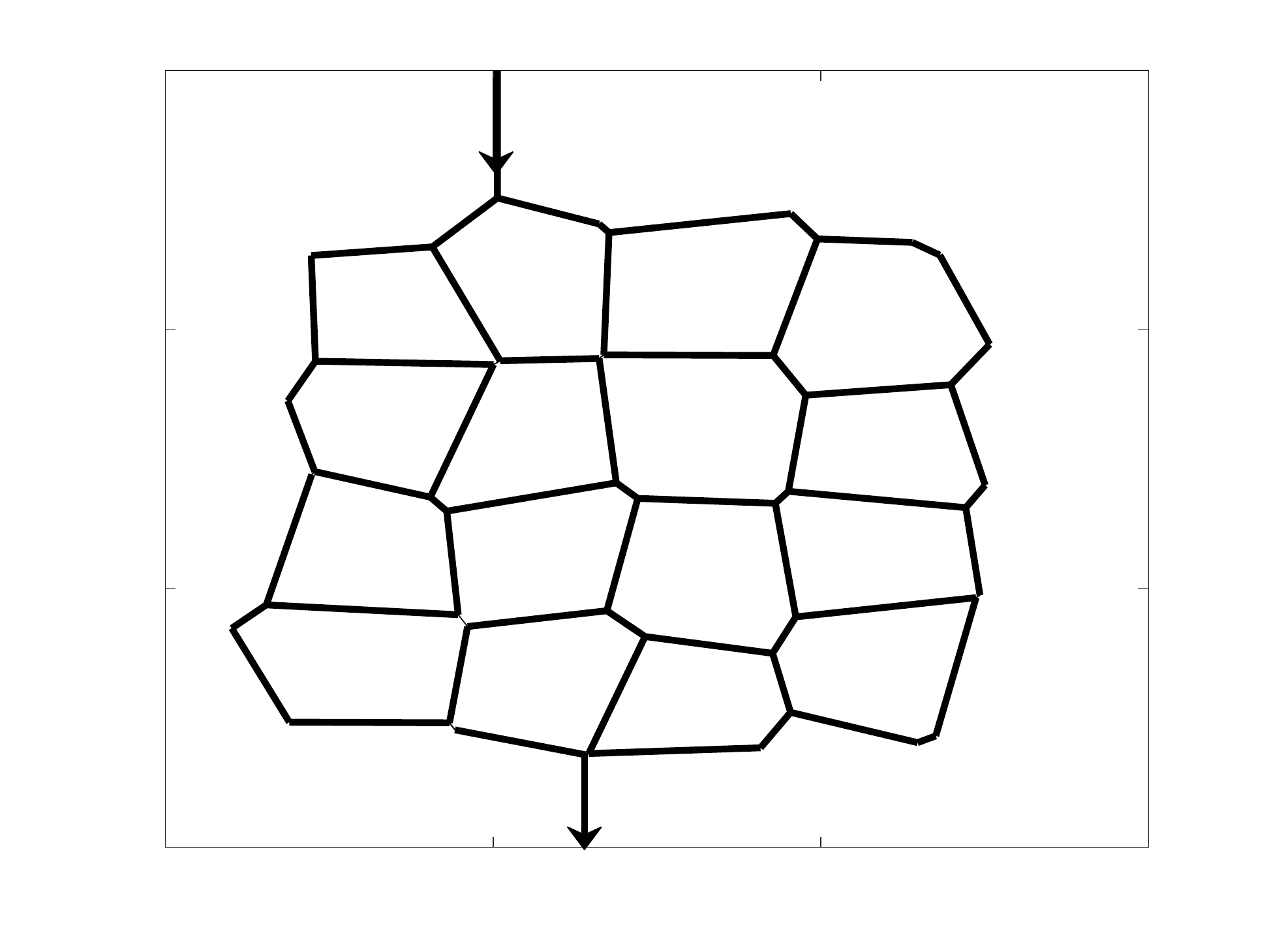}
\caption{Voronoi network}
 \label{vor}
\end{figure}

We numerically computed $\Delta \Gamma_{ij}$ for a class of networks,
as shown in Figure \ref{vor}A. This graph 
corresponds to a Voronoi
diagram with 4 $\times$ 4 cells. To generate a Voronoi diagram with M
$\times$ N  cells, we  choose random points
$$(x_j,y_k) \in \{ (x,y): j-1 < x < j, ~ k-1 < y < k \}$$
where $1 \leq j \leq M$ and $1 \leq k \leq N$. These points are then
used to generate a Voronoi diagram within Matlab. We next remove those
edges that intersect the region outside the rectangle $\{ (x,y): 0 < x
<  M, ~ 0 < y <  N \}$. Finally, we add incoming and
outgoing nodes and edges as follows. Suppose that the nodes of the diagram
constructed so far are at $\{ (x_i, y_k) \}$. Choose  nodes $min \anda
max$ so that the remaining nodes satisfy $y_{min} < y_k < y_{max}$. The incoming
edge then connects the point $(x_{max}, N+1)$ with the node at
$(x_{max},y_{max})$.
The outgoing
edge  connects the point $(x_{min}, 0)$ with the node at
$(x_{min},y_{min}).$

For each $M \anda N$ with $1 \leq M, \, N \leq 5$, we computed
$\Delta \Gamma_{ij}$ for each edge in 1000 randomly chosen Voronoi
networks of size $M \times N$. In every case, $\Delta \Gamma_{ij} > 0$.
\section{Grid Networks}

\begin{figure}
\centering
\includegraphics[width=5.5in,height=3in]{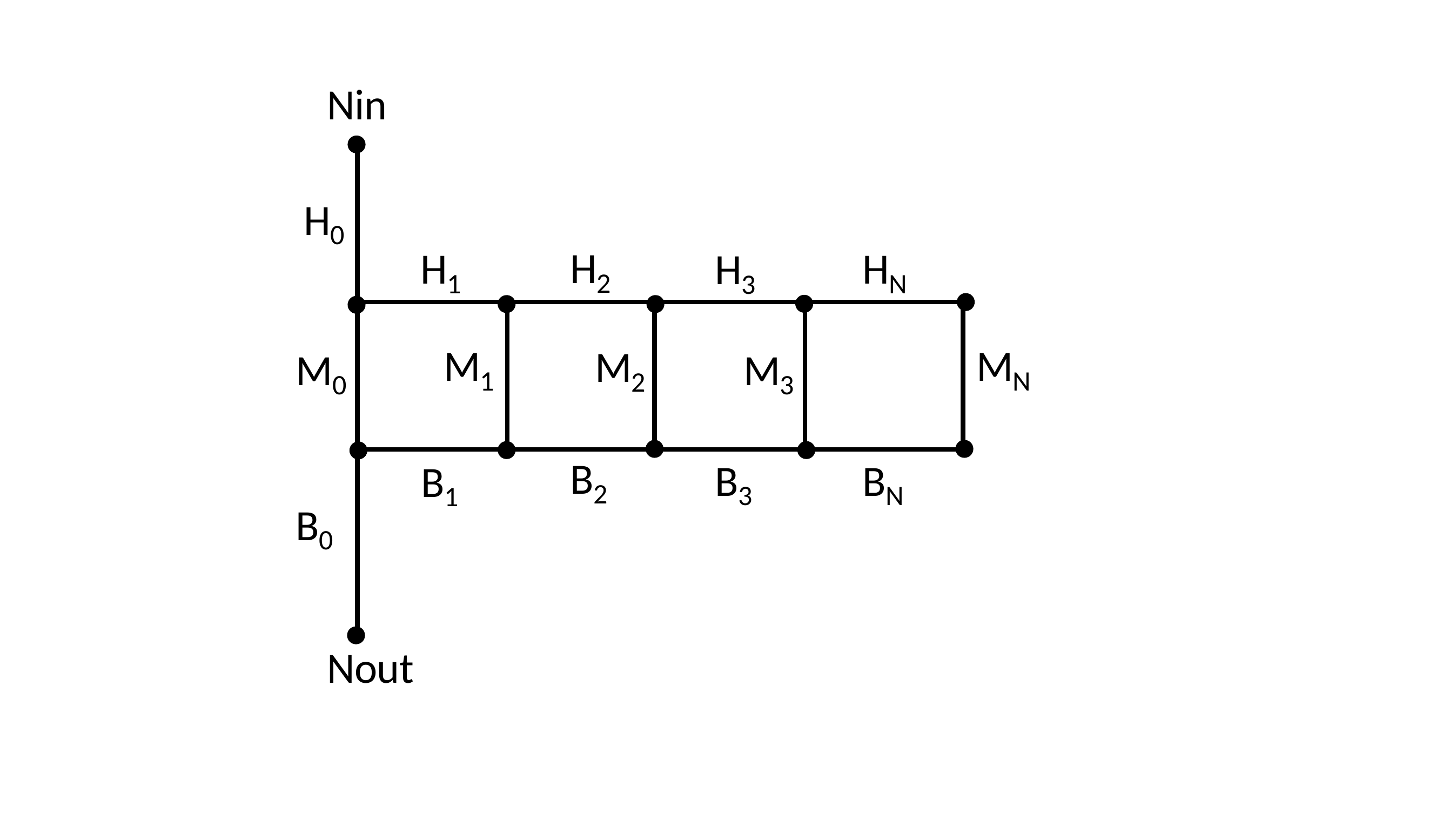}
\caption{The network ${\cal G}_N$}
\label{schema}
\end{figure}

We now consider the graph shown in Figure \ref{schema}B, which we
denote as ${\cal G}_N$. We will rigoursly prove that the Main Result
is valid for all  $N \geq 0$.

\subsection{Trees and Forests}

We  assume, without loss of generality, $P_{in} = 1
\anda P_{out}=0$. If, in addition, each conductance $\alpha_{ij}=1$,
then we can rewrite (\ref{dgamma1}) as
\be \label{dgamma}
\frac{\partial^2 \, \Gamma_{ij}}{\partial \, \alpha^2_{rs}} &=& 
2 \left( 1 - 
  \frac{ D^{i0}_{rs}-D^{j0}_{rs}} {\delta_i-\delta_j} \right)
\left( \frac{\delta}{\delta_i - \delta_j} \right)
\left( \frac{A^0_{rs}}{\delta}-
\frac{   D^{i0}_{rs}-D^{j0}_{rs} }
{\delta_i-\delta_j}
\right).
\ee
One can  interpret each term in (\ref{dgamma}) as
the number of trees and forests of some graph \cite{wkchen}.
If $\gamma \anda \beta$ are any distinct edges, let

\begin{itemize}
\item ${\cal T}_N ~=$  the number of trees of ${\cal G}_N$.
\item $\sigma_N ~=$ the number of 2-forests of ${\cal G}_N$ so that
  $N_{in}
  \anda N_{out}$ are
in different trees.
\item $\sigma_N(\gamma) ~=$ the number  of 2-forests of  ${\cal
  G_N} \backslash \gamma$ so that $N_{in} \anda N_{out}$ are in different trees.
\item ${\cal P}_N(\beta)  ~=$ the number of trees in ${\cal G}_N$ so
  that the unique path from $N_{in}$ to $N_{out}$ passes through $\beta$.
\item ${\cal P}_N(\gamma, \beta) ~=$ the number trees in ${\cal  G}_N \backslash \gamma$ so
  that the unique path from $N_{in}$ to $N_{out}$ passes through $\beta$.
\end{itemize}
If $\gamma \neq \beta$ correspond to the edges $e_{rs} \anda e_{ij}$,
respectively,
then for each $N$, 
\be
\delta &=& \sigma_N,  ~~~~~~~~ A^0_{rs} \, = \, \sigma_N(\gamma), \nn \\
\delta_i-\delta_j &=& {\cal P}_N(\beta), ~~~ D^{i0}_{rs}-D^{j0}_{rs}
\, = \,  {\cal P}_N(\gamma, \beta). \nn
\ee
Hence, we can rewrite (\ref{dgamma}) as
\be
\label{dq2}
\frac{\partial^2 \, \Gamma_{\beta}}{\partial \, \gamma^2} &=&
2 \left( 1 -\frac{ {\cal P}_N (\gamma,\beta)}{ {\cal P}_N(\beta)}
\right)
\left( \frac{\sigma_N}{{\cal P}_N(\beta)} \right)
\left( \frac{ \sigma_N(\gamma) }{\sigma_N} -
\frac{ {\cal P}_N(\gamma,\beta)}{ {\cal P}_N(\beta)} \right).
\ee
In a similar manner, we can rewrite (\ref{dgamma3}) and
(\ref{dgamma5}) as (\ref{dq2}). 
If $\gamma = \beta$, then we can rewrite (\ref{dgamma2}), 
(\ref{dgamma4}) and (\ref{dgamma6}) as
\be \label{dq2a}
\frac{\partial^2 \, \Gamma_{\beta}}{\partial \, \gamma^2} &=&
2 \, \frac{\sigma_N(\beta) }{ {\cal P}_N(\beta) }. 
\ee

\subsection{Formulas}

\begin{figure}
\centering
\includegraphics[width=6in,height=4in]{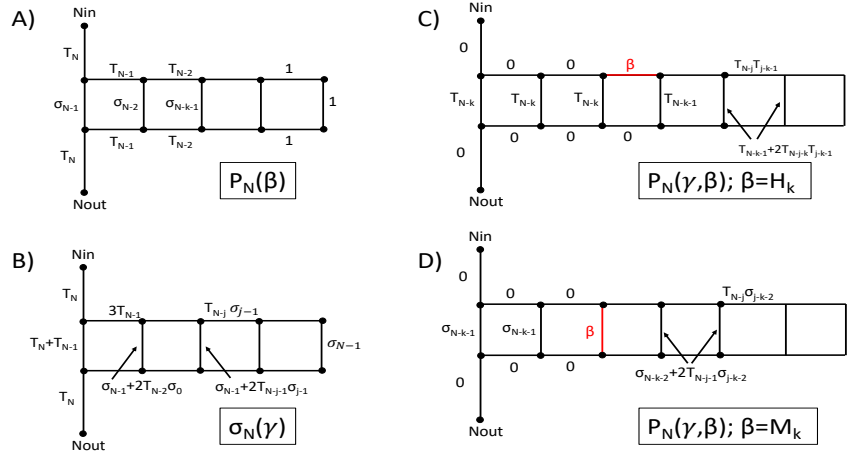}
\caption{ A) $ {\cal P}_N(\beta)$.  B)  $\sigma_N(\gamma)$. 
C) ${\cal P}_N(\gamma,\beta); ~ \beta=H_K$.  D) ${\cal
  P}_N(\gamma,\beta); ~ \beta=M_K$.}
 \label{formulas}
\end{figure}

The following formulas are
derived in Section \ref{derivation}. These results are illustrated in
Figure \ref{formulas}.

\medskip
  
\noindent  (F1) ~ $T_{N+1} = 4 \, T_N - T_{N-1}; ~~~~ T_0 = 1, ~T_1=4.$
\bigskip 

\noindent  (F2) ~ $\sigma_{N+1} = 4 \, \sigma_N - \sigma_{N-1}; ~~~~ \sigma_0 = 3,
    ~\sigma_1=11.$
\bigskip

\noindent  (F3) ~ \be
 {\cal P}_N(\beta) = & T_{N-k} & {\rm if} ~ \beta = H_k ~{\rm or}~  B_k
 \nn \\
 & \sigma_{N-k-1} &  {\rm if} ~ \beta = M_k.
 \nn
 \ee
 \bigskip

\noindent (F4) ~ \be
\sigma_N(\gamma) =
& T_{N-j} \, \sigma_{j-1}  & {\rm if}~ \gamma = H_j ~{\rm or}~ B_j  
\nn \\
& \sigma_{N-1} + 2 T_{N-j-1}\sigma_{j-1} & {\rm if}~ \gamma = M_j. \nn
\ee
\bigskip

\noindent   (F5) ~ If $\beta=H_k$ or $B_k$ and $\gamma= M_j$, then
\be
{\cal P}_N(\gamma,\beta) =& T_{N-k} & {\rm if} ~ 0 \leq j \leq k-1
\nn \\
&T_{N-k-1} & {\rm if} ~ j = k \nn \\
& T_{N-k-1}+2 \, T_{N-j-k} \, T_{j-k-1} &  {\rm if} ~ k < j \leq N.
\nn 
\ee
\bigskip

\noindent  (F6) ~ If $\beta=H_k$ or $B_k$ and $\gamma= H_j$ or $B_j$, then
\be
{\cal P}_N(\gamma,\beta) =& 0 & {\rm if} ~ 0 \leq j \leq k
\nn \\
& T_{N-j} \, T_{j-k-1} & {\rm if} ~ k+1 \leq j \leq N. \nn
\ee
\bigskip

\noindent (F7) ~  If $\beta=M_k$ and $\gamma= M_j$, then
\be
{\cal P}_N(\gamma,\beta) =&  \sigma_{N-k-1}  & {\rm if} ~ 0 \leq j \leq k
\nn \\
& \sigma_{N-k-2} + 2T_{N-j-1} \sigma_{j-k-2} & {\rm if} ~ k+1 \leq j \leq N. \nn
\ee
\bigskip

\noindent (F8) ~  If $\beta=M_k$ and $\gamma= H_j$ or $B_j$, then
\be
{\cal P}_N(\gamma,\beta) =& 0 & {\rm if} ~ 0 \leq j \leq k
\nn \\
& T_{N-j} \, \sigma_{j-k-2} & {\rm if} ~ k+1 \leq  j \leq N.
\nn
\ee

\subsection{Some useful identities}

The following identities will be used throughout the analysis:
\bigskip

\noindent A1) ~ $T_{N+1} = T_N + \sigma_N$.
\bigskip

\noindent A2) ~ $\lim_{N \rightarrow \infty} \frac{T_N}{T_{N+1}} ~=~
  \lim_{N \rightarrow \infty} \frac{\sigma_N}{\sigma_{N+1}} ~=~
  x_* \equiv 2-\sqrt{3}.$ 
  \bigskip
  
\noindent A3) ~ $\frac{T_N}{T_{N+1}}\, < \, x_* \, < \,
    \frac{\sigma_N}{\sigma_{N+1}} \, < \, 3/11$. Moreover, $\frac{T_N}{T_{N+1}} \anda
    \frac{\sigma_N}{\sigma_{N+1}}$
    are increasing and
decreasing functions of  $N$, respectively.,
    \bigskip

\noindent A4)  ~  $\lim_{N\rightarrow \infty} \frac{T_N}{\sigma_N} ~=~
  \frac{x_*}{1-x_*}$.
  \bigskip

\noindent A5) ~ If $0 \leq k < j \leq N$, then
$
  \frac{T_{j-k-1}}{T_{N-k}} ~<~ \frac{\sigma_{j-1}}{\sigma_N}. 
 $
  \bigskip

  \noindent A6) ~ $\sigma_k \, x_*^k \, > \, 2.5$ for all $k \geq 0$.
  \bigskip
  
\noindent  A7) ~ $ {\cal P}(\gamma,\beta) \, \leq \, {\cal
      P}(\beta)$ for every pair of edges $\beta \anda \gamma$.
\bigskip

The proof of A1) is by
induction. It is true when $N=0$ since $T_0=1, \, T_1=4 \anda \sigma_0=3$.
Suppose that 1 is true up to
some $N$. Then using (F1) and (F2),
\be
T_{N+1} &=& T_N+\sigma_N ~\Rightarrow~
4T_{N+1} = 4T_N+4\sigma_N \nn \\
&\Rightarrow & 
T_{N+2}+T_N = T_{N+1} + T_{N-1} + \sigma_{N+1} + \sigma_{N_1} \nn \\
& \Rightarrow& T_{N+2} =  T_{N+1} + \sigma_{N+1.} \nn
\ee
\medskip

To prove A2), let $x_N = \frac{T_N}{T_{N+1}}$ or
$\frac{\sigma_N}{\sigma_{N+1}}$.
Then (F1) implies that $x_{N+1} =
1/(4-x_{N})$. As $N \rightarrow \infty$, $x_N$ approaches the stable fixed
point of this map. This fixed point satisfies $x_*^2 - 4x_*+1=0$. That
is,  $x_* =  2-\sqrt{3}$.
\medskip

To prove A3), let $x_N = \frac{T_N}{T_{N+1}}$. Then
$x_0 = 1/4 < x_*$. Moreover, if $x_N < x_*$,  then
$$
x_{N+1} ~=~ 4 - \frac{1}{x_N} \,> \, x_N
$$ and
$$x_{N+1} ~=~ 4 - \frac{1}{x_N} \,< \, 4 - \frac{1}{x_*} \, = \, x_*
$$
A similar argument hold for $\frac{\sigma_N}{\sigma_{N+1}}$. Moreover,
$\frac{\sigma_N}{\sigma_{N+1}} \, < \, \frac{\sigma_0}{\sigma_1} = 3/11$.
\medskip

To prove A4), note that
$$
\frac{T_N}{\sigma_N} ~=~ \frac{T_N}{T_{N+1}-T_N} ~=~
\frac{T_N/T_{N+1}}{1-T_N/T_{N+1}}  ~\rightarrow \frac{x_*}{1-x_*}
$$
as $N \rightarrow \infty$.
\medskip

A5) follows from A3) because
\be
\frac{T_{j-k-1}}{T_{N-k}} &=& \frac{T_{j-k-1}}{T_{j-k}}
\frac{T_{j-k}}{T_{j-k+1}} \dots \frac{T_{N-k-1}}{T_{N-k}} \nn \\ \nn \\
&<& \frac{T_{j-1}}{T_{j}} \frac{T_j}{T_{j+1}} \dots
\frac{T_{N-1}}{T_N} ~~~ <~~
 \frac{\sigma_{j-1}}{\sigma_{j}} \frac{\sigma_j}{\sigma_{j+1}} \dots
\frac{\sigma_{N-1}}{\sigma_N} \nn \\ \nn  \\
&=& \frac{\sigma_{j-1}}{\sigma_N} \nn
\ee 
\medskip

To prove A6), let $B_k = \sigma_k \, x_*^k$. Then
\be
B_{k+1} &=& \sigma_{k+1} \, x_*^{k+1} ~=~
4 \, \sigma_k \,_*^{k+1} \, - \, \sigma_{k-1} \, x_*^{k+1} \nn \\
&=& 4 \, x_* \, B_k \, - \, x_*^2 \, B_{k-1} \nn \\
&=&  4 \, x_* \, (B_k \, - \, B_{k-1}) \, + \, B_{k-1}. \nn
\ee
Since $x_*^2 -4x_* -1 = 0$. Hence,
\be
\frac{ B_{k+1} \, - \, B_{k-1}}{B_k \, - \, B_{k-1}} &=& 4x_* \nn
\ee
and, therefore,
\be
\frac{B_{k+1} \, - \, B_k}{B_k - B_{k-1}} &=& 4x_* \, - \, 1 ~ = ~
x_*^2. \nn
\ee
Hence, for all $k \geq 0$.
\be
B_{k+1} &=& B_k \, + \, x_*^2 \, (B_k \, - \, B_{k-1}). \nn
\ee
Since $B_0 = 3 \anda B_1 = 11x_* < B_0 $,
it follows that 
\be
B_k &=& B_1 \, + \,  (B_1 \, - \, B_0) \, \sum_{i=1}^{2(k-1)} x_*^{2i}
\nn \\
&>& B_1 \, + \, (B_1 \, - \, B_0) \, \sum_{i=1}^{\infty} x_*^{2i} \nn \\
&=& B_1 \, + \, (B_1 \, - \, B_0) \left( \frac{x_*^2}{1-x_*^2}
\right)   \nn \\
& \approx & 2.9434 ~>~ 2.5. \nn
\ee
\medskip

Finally, A7) is true because every tree in ${\cal G} \backslash \gamma$
such that the unique path from $N_{in}$ to $N_{out}$ passes through $\beta$
is also a tree in ${\cal G}$ with the same property.

\subsection{Derivatives}

We now use (\ref{dq2}) and (\ref{dq2a}) to compute second derivatives. There are many
cases to consider.

\begin{enumerate}

\item $\beta= H_k$ or $B_k$  and $\gamma = H_j, ~0 \leq j<k$.
  Then
  $${\cal P}_N(\beta)=T_{N-k}, ~ ~~\sigma_N(\gamma) = T_{N-j} \,
  \sigma_{j-1},
  ~~~ {\cal  P}(\gamma,\beta)=0.$$
Hence,
\be
\frac{\partial^2 \, \Gamma_{\beta}}{\partial \, \gamma^2}  &=& 
2 \, \frac{ T_{N-j} ~ \sigma_{j-1} }{T_{N-k}} ~>~ 0. \nn 
\ee

\item $\beta= H_k$ or $B_k$ and $\gamma = H_j, ~k<j \leq N$.
  Then
  $${\cal P}_N(\beta)=T_{N-k}, ~~~ \sigma_N(\gamma)=T_{N-j} \,
  \sigma_{j-1}, 
  ~~~ {\cal  P}_N(\gamma,\beta) = T_{N-j} \, T_{j-k-1}.$$
Hence, using the Identities A5 and A7,
\be
\frac{\partial^2 \, \Gamma_{\beta}}{\partial \, \gamma^2}  
&=& 2 
\left( 1 - \frac{ {\cal P}_N(\gamma,\beta)} { {\cal P}_N(\beta) }
\right)
\, \left( \frac{\sigma_N \, T_{N-j}}{T_{N-k}} \right)
\left( \frac{\sigma_{j-1}}{\sigma_N} - \frac{T_{j-k-1}}{T_{N-k}}
\right). \nn \\
&>& 0. \nn
\ee

\item  $\beta= H_k$ or $B_k$ and $\gamma = M_j, ~ 0
  \leq j \leq k-1$. Then
   ${\cal P}_N(\beta) \, = \, {\cal P}_N(\gamma,\beta) =T_{N-k}$.
Hence, 
\be
\frac{\partial^2 \, \Gamma_{\beta}}{\partial \, \gamma^2}  &=& 
0 \nn
\ee

\item  $\beta= H_k$ or $B_k$ and $\gamma = M_j, ~ k+1
  \leq j \leq N$. Then
  $${\cal P}_N(\beta) = T_{N-k}, ~~~ 
 {\cal  P}(\gamma,\beta) =T_{N-k-1}+2 \, T_{N-j-k} \, T_{j-k-1}, ~~~
 \sigma_N(\gamma) = \sigma_{N-1}+ 2 \, T_{N-j-1}\sigma_{j-1}. $$
Hence, using Identites A5 and A7,
\be
\frac{\partial^2 \, \Gamma_{\beta}}{\partial \, \gamma^2}  &=&
\Lambda_1 \,  \Lambda_2 \nn
\ee
where
\be
\Lambda_1 &=& 2 \, \left( \frac{\sigma_N}{T_{N-k}} \right) \,
\left( 1-\frac{ {\cal P}_N(\gamma,\beta)}{ {\cal P}_N(\beta)} \right)
~>~ 0
\nn
\ee
and
\be
\Lambda_2 &=& \frac{\sigma_{N-1}+ 2 \,
  T_{N-j-1}\sigma_{j-1}}{\sigma_N}
\, - \, \frac{ T_{N-k-1}+2 \, T_{N-j-k} \, T_{j-k-1}}{T_{N-k}}  \nn \\
\nn \\
&=& \left( \frac{\sigma_{N-1}} {\sigma_N} 
  - \frac{T_{N-k-1}}{T_{N-k}} \right) \, + \, 2 \, T_{N-j-1} \,
\left( \frac{ \sigma_{j-1}}{\sigma_N} - \frac{ T_{j-k-1}}{T_{N-k}}
\right) \nn \\ \nn \\
&>& 0. \nn
\ee

\item  $\beta= M_k$  and $\gamma = H_j$ or $B_j, 0 \leq j \leq k$.
  Then
  $${\cal P}_N(\beta) = \sigma_{N-k-1}, ~~~
  {\cal P}_N(\gamma,\beta) = 0, ~~~ \sigma_N(\gamma) = T_{N-j} \, \sigma_{j-1}.$$
Hence,
\be
\frac{\partial^2 \, \Gamma_{\beta}}{\partial \, \gamma^2}  &=&
2 \, \frac{ T_{N-j} \, \sigma_{j-1}}{\sigma_{N-k-1}}. \nn
\ee
Note that $T_{N-j} \, \sigma_{j-1}$ is a decreasing function of
$j$. This is because, from Identity A3,
$$
\frac{\sigma_{j-1}}{\sigma_{j}} ~>~ x_* ~>~
\frac{T_{N-j-1}}{T_{N-j}}.
$$
Hence,
\be
\frac{\partial^2 \, \Gamma_{\beta}}{\partial \, \gamma^2}  &>&
2 \, \frac{ T_{N-k} \, \sigma_{k-1}}{\sigma_{N-k-1}}. \nn
\ee

\item  $\beta= M_k$  and $\gamma = H_j$ or $B_j,~ k < j \leq N$.
  Then
  $${\cal P}_N(\beta) = \sigma_{N-k-1} ~~~
  {\cal P}_N(\gamma,\beta) =  T_{N-j} \, \sigma_{j-k-2}, ~~~
  \sigma_N(\gamma) = T_{N-j} \, \sigma_{j-1}.$$
Hence,
\be
\frac{\partial^2 \, \Gamma_{\beta}}{\partial \, \gamma^2}  &=&
2 \, \Lambda_1 \, \Lambda_2 \, \Lambda_3  \, \Lambda_4 \nn
\ee
where
\be
\Lambda_1 &=& 1-\frac{{\cal P}_N(\gamma,\beta)}{ {\cal P}_N(\beta)};
~~~~~
\Lambda_2 =
\frac{ T_{N-j} \, \sigma_{j-1}}{\sigma_{N-k-1}}; \nn \\
\Lambda_3
&=&  \frac{\sigma_N}{\sigma_{N-k-1}}; ~~~~~~~~~~~~
\Lambda_4 = \frac{\sigma_{N-k-1}}{\sigma_N} -
  \frac{\sigma_{j-k-2}}{\sigma_{j-1}}.  \nn
  \ee
 From Identity A7 , $0 < \Lambda_1 < 1$. 
It was shown earlier that $T_{N-j} \, \sigma_{j-1}$ is a decreasing function of
$j$. Since $j > k$,
$$
0 ~<~ \Lambda_2 ~<~ \frac{T_{N-k} \, \sigma_{k-1}}{\sigma_{N-k-1}}. 
$$
Using Identity A3,
\be
\Lambda_3 &<& \frac{\sigma_N}{\sigma_{N-1}} \,
\frac{\sigma_{N-1}}{\sigma_{N-2}}
\dots \frac{\sigma_{N-k-2}}{\sigma_{N-k-1}} \nn \\ \nn \\
&<& 
x_*^{-(k+1)}. \nn
\ee
Now consider $\Lambda_4$.  Let
$$A_j ~=~ \frac{\sigma_{j-k-1}}{\sigma_j}.$$
Then
\be
A_{j+1} &=& \frac{\sigma_{j-k}}{\sigma_{j+1}} ~=~
\frac{4 \, \sigma_{j-k-1}-\sigma_{j-k-2}}{4 \, \sigma_j -
  \sigma_{j-1}} \nn \\
&=& \frac{A_j - x_j \, A_{j-1}}{ 1 - x_j} \nn \\
& =& A_j + \lambda_j \, (A_j - A_{j-1}) \nn
\ee
where
\be
x_j &=& \frac{\sigma_{j-1}}{4 \, \sigma_j}
~~ \anda ~~
\lambda_j \, = \, \frac{x_j}{1-x_j} \, = \,
\frac{\sigma_{j-1}}{\sigma_{j+1}}. \nn
\ee
Let \,
$\epsilon = A_{k+1} \, - \, A_k.$
Then, for $n \geq 1$,
\be
A_{k+n} &=& A_{k+1} \, + \, \epsilon \, \sum_{i=1}^{n-1}
\prod_{y=1}^i
\lambda_{k+y}. \nn
\ee
Hence,
\be
A_N \, - \, A_{j-1} &=&
\epsilon \, \left(
  1 + \sum_{i=1}^{N-j-1} \prod_{y=1}^{i} \lambda_{j+y} \right)
 \prod_{y = k+1}^{j-1} \lambda_y.
  \nn
  \ee
From Identity A3,  $\lambda_j < y_*^2$ where $y_* = 3/11$. Moreover,
\be
\epsilon  &=& \ A_{k+1} \, - \,  A_k
~=~  \frac{3}{\sigma_{k+1}} -
\frac{1}{\sigma_k}  \nn \\
&=&  \frac{3 \sigma_k - \sigma_{k+1}}{\sigma_k \, \sigma_{k+1}} 
~=~ \frac{\sigma_{k-1} - \sigma_k}{\sigma_k \, \sigma_{k+1}} 
~ < ~0. \nn
\ee
This implies that 
~$\frac{\partial^2 \, \Gamma_{\beta}}{\partial \, \gamma^2}  \, < \,
0 $ \, and , \, $|\epsilon| \, < \, \frac{1}{\sigma_{k+1}}. $ \, 
It follows that
\be \label{A4}
| \Lambda_4 | &=&
| A_N - A_{j-1} | ~<~
|\epsilon| \, y_*^{2 \, (j-k+1)} \, \sum_{i=0}^{\infty} y_*^{2i}  \nn \\
& < &
\frac{y_*^{2 \, (j- k-1)}}{\sigma_{k+1} \, (1-y_*^2)}. 
\ee

We have therefore shown that
\be \label{deriv6}
\left| \frac{\partial^2 \, \Gamma_{\beta}}{\partial \, \gamma^2}  \right|
&<& 2 \, \left( \frac{T_{N-k} \, \sigma_{k-1}}{\sigma_{N-k-1}} \right)
\left( \frac{1}{x_*^{k+1} \, \sigma_{k+1}} \right)
\left( \frac{y_*^{2 \, (j- k-1)}}{ 1-y_*^2} \right) \nn \\ \nn \\
& < & 
.8 \, \left( \frac{T_{N-k} \, \sigma_{k-1}}{\sigma_{N-k-1}} \right)
\left( \frac{y_*^{2 \, (j-k-1)}}{1-y_*^2} \right). \nn
\ee
Here we used Identity A6.

\medskip

\item  $\beta= M_k$  and $\gamma = M_j, ~ 0 \leq j < k$.
  Then ${\cal P}_N(\beta) =  {\cal P}_N(\gamma,\beta) = \sigma_{N-k-1}.$
Hence,
\be
\frac{\partial^2 \, \Gamma_{\beta}}{\partial \, \gamma^2}  &=& 0. \nn
\ee

\item  $\beta= M_k$  and $\gamma = M_j, ~ k+1 \leq j \leq N$.
  Then
  $$ {\cal P}_N(\gamma,\beta) = \sigma_{N-k-2}+2T_{N-j-1} \,
  \sigma_{j-k-2},
  ~~~ {\cal P}_N(\beta) = \sigma_{N-k-1}, ~~~
   \sigma_N(\gamma) = \sigma_{N-1} + 2 \, T_{N-j-1} \, \sigma_{j-1}. $$
   Hence,
   \be
\frac{\partial^2 \, \Gamma_{\beta}}{\partial \, \gamma^2}  &=&
\Lambda_5 \, \Lambda_6 \nn
\ee
where, using Identity A3,
\be
\Lambda_5 &=&
2 \,\left( \frac{\sigma_N}{\sigma_{N-k-1}} \right)
\left( 1 - \frac{ {\cal P}(\gamma,\beta) } { {\cal P}(\beta)} \right)
\, < \, 2 x_*^{-(k+1)} \nn
\ee
and
\be
\Lambda_6 &=&
  \frac{\sigma_{N-1} + 2 \, T_{N-j-1} \, \sigma_{j-1}}{\sigma_N} -
\frac{\sigma_{N-k-2}+2T_{N-j-1} \,
  \sigma_{j-k-2}}{\sigma_{N-k-1}} \nn \\ \nn \\
&=& \left( \frac{\sigma_{N-1}}{\sigma_N} -
  \frac{\sigma_{N-k-2}}{\sigma_{N-k-1}} \right)
+ 2 \, T_{N-j-1} \left( \frac{\sigma_{j-1}}{\sigma_N} -
  \frac{\sigma_{j-k-2}}{\sigma_{N-k-1}}  \right). \nn \\ \nn \\
&>& 2 \, T_{N-j-1} \left( \frac{\sigma_{j-1}}{\sigma_N} -
  \frac{\sigma_{j-k-2}}{\sigma_{N-k-1}}  \right) \nn \\ \nn \\
&=&
 2 \, \left( \frac{ T_{N-j-1} \, \sigma_{j-1}}{\sigma_{N-k-1}} \right)
 \, \Lambda_4. \nn
 \ee
 Since $\Lambda_4 < 0$, this implies that \,
 $\frac{\partial^2 \, \Gamma_{\beta}}{\partial \, \gamma^2} \, < \, 0$.
 Moreover, using (\ref{A4}), 
\be
| \Lambda_6 | &<& 2 \,
\left( \frac{ T_{N-k-1} \, \sigma_{k-1}}{\sigma_{N-k-1}} \right) \nn
\left( \frac{ y_*^{2 \, (j-k-1)}}{\sigma_{k+1} \, (1-y_*^2)} \right)
\ee
Hence, using Identity A6,
\be
\left| \frac{\partial^2 \, \Gamma_{\beta}}{\partial \, \gamma^2} \right|
&<& 4 \, 
\left( \frac{ T_{N-k-1} \, \sigma_{k-1}}{\sigma_{N-k-1}} \right)
\left( \frac{y_*^{2 \, (j-k-1)}}{1-y_*^2} \right)
\left( \frac{1}{y_*^{k+1} \, \sigma_{k+1}} \right) \nn \\ \nn \\
&<& 1.6 \, \left( \frac{ T_{N-k-1} \, \sigma_{k-1}}{\sigma_{N-k-1}} \right) \nn
\left( \frac{y_*^{2 \, (j-k-1)}}{1-y_*^2}  \right).
\ee

\item $\beta = H_k$ or $B_k \anda \gamma = \beta$. Then ${\cal P}_N(\beta) =
  T_{N-k} \anda \sigma_N(\gamma) = T_{N-k} \, \sigma_{k-1}$. Hence,
\be
\frac{\partial^2 \, \Gamma_{\beta}}{\partial \, \gamma^2} &=&
2 \, \frac{\sigma_N(\beta)}{ {\cal P}_N(\beta)} ~=~
2 \,  \sigma_{k-1}. \nn
\ee

 \item $\beta = M_k \anda \gamma = \beta$. Then ${\cal
     P}_N(\beta) = \sigma_{N-k-1} \anda \sigma_N(\gamma) = \sigma_{N-1}
  + 2 T_{N-k-1} \sigma_{k-1}$. Hence,
   \be
   \frac{\partial^2 \, \Gamma_{\beta}}{\partial \, \gamma^2}  &=&
   2 \,  
   \frac{\sigma_{N-1}
     + 2 T_{N-k-1} \sigma_{k-1}} { \sigma_{N-k-1}} ~ > ~ 2.
 \nn
   \ee

 \end{enumerate}
 
   \subsection{The Laplacian}

   We have demonstrated that if $\beta = H_k$ or $B_k$, then for each $\gamma$,
     ~$   \frac{\partial^2 \, \Gamma_{\beta}}{\partial \, \gamma^2}
     \, > \, 0$.
Hence, 
$   \Delta \, \Gamma_{\beta} \, > \,  0. $
   \medskip

   Now supposse that $\beta = M_k$. We have demonstrated that
   \be
   \frac{\partial^2 \, \Gamma_{\beta}}{\partial \, \gamma^2}  &>& 0
   ~~~ {\rm if} ~~ \gamma = H_k, \, B_k ~{\rm or}~ M_k ~\anda~ 0 \leq j
   \leq k \nn \\
   &<& 0    ~~~ {\rm if} ~~ \gamma = H_k, \, B_k ~{\rm or}~ M_k ~\anda~ k < j
             \leq N. \nn
   \ee
   From the above analysis, and recalling that $y_* = 3/11$,
   \be
\Delta \, \Gamma_\beta
 &>&   \frac{T_{N-k} \, \sigma_{k-1}}{\sigma_{N-k-1}}
 \left( 2(k+1) \, + \, 2 \, - \,  \left( \frac{3,2}{1-y_*^2} \right)
   \sum_{j = k+1}^N y_*^{2 \, (j-k-1)} \right) \nn \\
&>&  \frac{T_{N-k}}{\sigma_{N-k-1}}
\left( 4 \, - \left( \frac{3.2}{1-y_*^2} \right)
  \sum_{j=0}^{\infty} y_*^{2j} \right) \nn \\
&=&  \frac{T_{N-k}}{\sigma_{N-k-1}}
\left( 4  \, - \, \frac{3.2}{(1-y_*^2)^2} \right) ~ > ~ 0. \nn
\ee
   
\section{Derivation of the Formulas} \label{derivation}

\noindent (F1) ~A proof of this result is given in \cite{Desjarlais,
  Raff}.
Clearly, $T_0 = 1 \anda T_1 = 4$. As shown in
Figure \ref{trees}A, there are 3 ways to extend each tree in ${\cal G}_N$ to
obtain a tree in ${\cal G}_{N+1}$.  If a tree in
${\cal G}_{N}$ contains the edge $M_N$, then we can obtain another tree
in ${\cal G}_{N+1}$ by removing this edge and adding the edges
$H_{N+1}, B_{N+1} \anda M_{N+1}$, as shown Figure \ref{trees}B.
It is not hard to show that the number of trees in ${\cal G}_N$ that
do not contain $M_N$ is precisely $T_{N-1}$.
\medskip

\noindent (F2) ~ This follows from an argument almost identical to
that for (F1).
\medskip

\noindent (F3) ~ If $\beta = H_k$ or $B_k$, then every path that goes from $N_{in}$ to
$N_{out}$ and which passes through $\beta$ must contain the
edges $H_i \anda B_i$ with $i \leq k$ (solid lines in Figure \ref{pnbeta}A).
 Every element of ${\cal P}_N(\beta)$
 is obtained by adding to these edges a tree for the graph
with edges $H_i, B_i \anda M_i$ with $i \geq k$ (shaded region in
Figure \ref{pnbeta}A). 
The number of such trees is $T_{N-k}$.

   If $\beta = M_k$, then every path that goes from $N_{in}$ to
  $N_{out}$ and which passes through $\beta$ must contain the edges $H_i
 \anda B_i, i \leq k,$ and $M_k$ (solid lines in Figure \ref{pnbeta}B).
 Every element of ${\cal P}_N(\beta)$
  is obtained by adding to these edges a 2-forest 
  for the graph with edges $H_i, B_i \anda M_i$ with $i \geq k$. 
The nodes  corresponding to the terminal ends of $\beta$ must be in
different trees.
The number of such forests is $\sigma_{N-k-1}$.
  \medskip

  \noindent (F4) ~  Suppose that $\gamma = H_j$. Then every element of
  $\sigma_N(\gamma)$ is of the form ${\cal F} \, \cup \, {\cal T} \,
  \cup \, B_j$ where
 ${\cal F}$ is
  any 2-forest for the graph with edges  $H_k, B_k \anda M_k$ with $k
  < j$ (blue region in Figure \ref{sigman}A) and 
 ${\cal T}$ is any tree for
  the graph with edges $H_k, B_k \anda M_k$ with $k \geq j$
  (grey region in Figure \ref{sigman}A).
  
  If $\gamma = M_j$, then
  consider the graph  that does not contain $\gamma$; moreover, $H_j \cup H_{j+1} \anda B_j \cup
B_{j+1}$ are combined into single edges. The number of 2-forests for this graph is $\sigma_{N-1}$.
Let ${\cal F}$ be one such forest. If {${\cal F}$ contains both of the combined edges, then
it is also an element of $\sigma_N(\gamma)$. If ${\cal
  F}$ does not contain, $H_j \cup H_{j+1}$  (or $B_j \cup B_{j+1}$) then we obtain
an element of $\sigma_N(\gamma)$ by adding the edge $H_j$ (or $B_j$) to ${\cal
  F}$, as shown in Figure \ref{sigman}C. This demonstrates that the
number of forests in $\sigma_N(\gamma)$ that contain both $H_j \anda
B_j$ is $\sigma_{N-1}$. To obtain an element of
$\sigma_N(\gamma)$ that does not contain $H_j$ (or $B_j$), let ${\cal F}$ be
any 2-forest for the graph with edges $H_k, B_k \anda M_k$ with $k < j$
(blue region in Figure \ref{sigman}D)
  and let ${\cal T}$ be any tree for 
  the graph with edges $H_k, B_k \anda M_k$ with $k \geq j$
(grey region in Figure \ref{sigman}). Then
  ${\cal  F} \, \cup  \, {\cal T} \, \cup \, B_j$ 
  (or   ${\cal  F} \, \cup  \, {\cal T} \, \cup \, H_j) \in \sigma_N(\gamma)$.
  \medskip
  
  \noindent (F5) ~ Suppose that $\beta = H_k \anda \gamma = M_j$. Then
  every path from $N_{in}$ to $N_{out}$ that passes through $\beta$ must
  contain the edges $E_k = \{H_i \anda B_i: 0 \leq i \leq k
  \}$ (green edges in Figure \ref{F5}).

  If $j
  < k$, let ${\cal T}$ be any tree in the graph with edges $H_i, B_i
  \anda M_n$ where $k+1 \leq i \leq N \anda  k \leq n \leq N$
(shaded region in Figure \ref{F5}A.) There are $T_{N-k}$ such
  trees and $E_k \, \cup \, {\cal F} \in {\cal P}_N(\gamma,\beta)$.

   If $j  = k$, let ${\cal T}$ be any tree in the graph with edges $H_i, B_i
  \anda M_i$ where $k+1 \leq i \leq N$. 
 (Shaded region in Figure \ref{F5}B.) There are $T_{N-k-1}$ such
  trees and $E_k \, \cup \, {\cal F} \in {\cal
    P}_N(\gamma,\beta)$.

 If $j > k$, then
  consider the graph  that does not contain  $\gamma$ and the edges
  $M_i, \, 0 \leq i <  k$. 
  Moreover, $H_j \cup H_{j+1} \anda B_j \cup B_{j+1}$
  are combined into single edges. The number of trees for this graph is $T_{N-k-1}$.
Let ${\cal T}$ be one such tree. If {${\cal T}$ contains both of the combined edges, then
$E_k \cup T$ is an element of ${\cal P}_N(\gamma,\beta)$. If ${\cal
  T}$
does not contain, $H_j \cup H_{j+1}$  (or $B_j \cup B_{j+1}$) then we obtain
an element of ${\cal P}_N(\gamma,\beta)$ by adding the edge $H_j$ (or $B_j$) to ${\cal
  T}$ and combining this with $E_k$. This demonstrates that the
number of trees in ${\cal P}_N(\gamma, \beta)$ that contain both $H_j \anda
B_j$ is $T_{N-k-1}$. To obtain an element of
${\cal P}_N(\gamma,\beta$ that does not contain $H_j$ (or $B_j$), let ${\cal T}$ be
any tree for the graph with edges $H_i, B_i \anda M_n$ with $j+1 \leq
i \anda j \leq n$
(blue region in Figure \ref{F5}C)
  and let ${\cal T}_0$ be any tree for 
  the graph with edges $H_i, B_i \anda M_n$ with $k+1 \leq i  \leq j-1
  \anda k \leq n \leq j-1$
(grey region in Figure \ref{F5}C). Then
  $E_k \cup {\cal  T}_0  \cup   {\cal T}  \cup  B_j   \cup 
  B_{j+1}  \cup H_{j+1}$ 
  (or   ${\cal  T}  \cup   {\cal T}_0  \cup  H_j  \cup   H_{j+1} 
  \cup  B_{j+1}) \in \sigma_N(\gamma)$.
  \medskip

  \noindent (F6) ~ Suppose that $\beta = H_k \anda \gamma = H_j$. If
  $j \leq k$, then there are no paths in ${\cal G}_N \backslash
  \gamma$ from $N_{in}$ to $N_{out}$ that passes through $\beta$. Hence,
  ${\cal P}_N(\gamma,\beta)=0$.

  If $j >k$, let $E_k$ be the set of edges defined above. Moreover,
  let
  ${\cal T}$ be any tree for the graph with edges $H_i, B_i, M_n$ with
  $k < i  < j \anda k \leq n \ <  j$ (grey region in Figure
  \ref{F6}B) and ${\cal T}_0$ be any tree for the graph with edges
  $H_i, B_i, M_n$ with $j+1 \leq i \leq N \anda j \leq n \leq N$ (blue
  region in Figure \ref{F6}B). Then $E_k \cup {\cal T} \cup {\cal
    T}_0 \in {\cal P}_N(\gamma,\beta)$.
  \medskip

  \noindent  (F7) and (F8) ~ The derivation is very similar to (F5)
  and (F6), respectively.

\begin{figure}
\centering
\includegraphics[width=4in,height=2in]{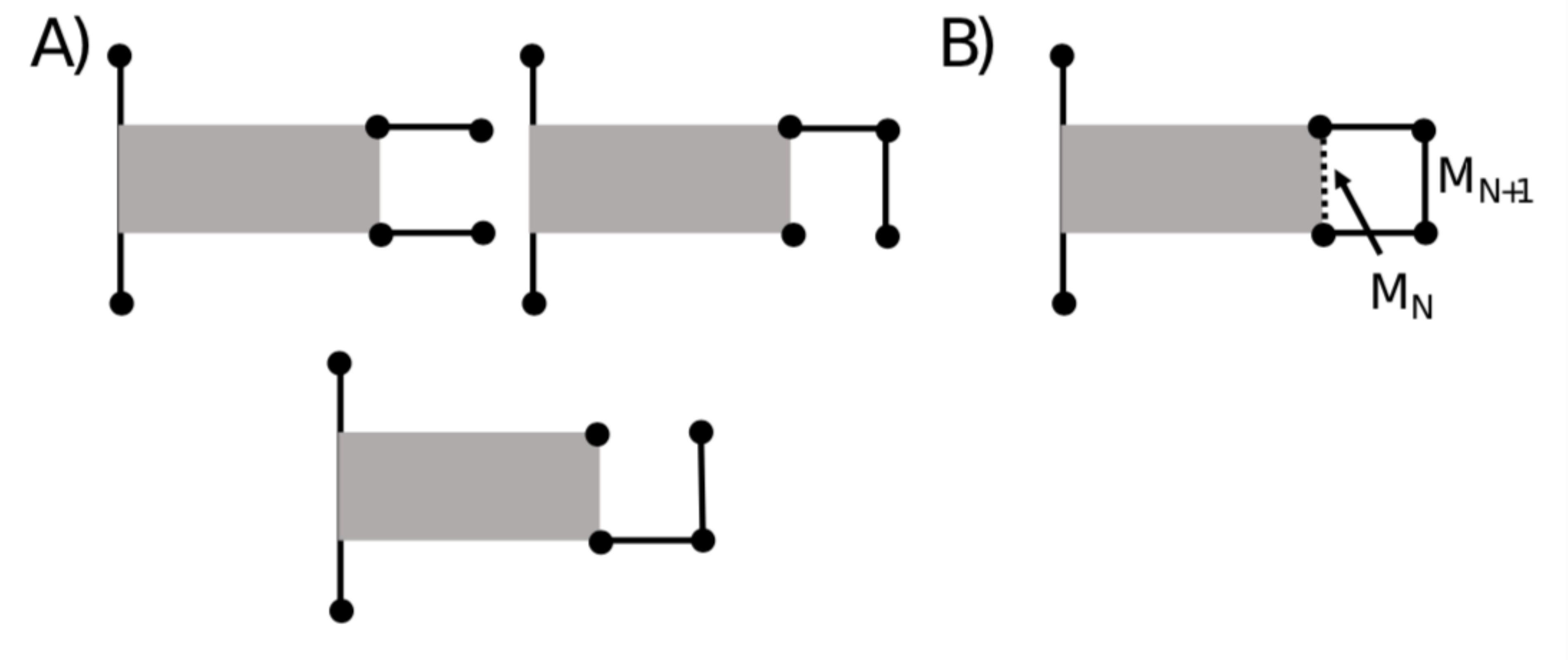}
\caption{Trees: A) There are 3 ways to extend each tree in ${\cal G}_N$ to
obtain a tree in ${\cal G}_{N+1}$. B)  If a tree in
${\cal G}_{N}$ contains the edge $M_N$, then we can obtain another tree
in ${\cal G}_{N+1}$ by removing this edge and adding the edges
$H_{N+1}, B_{N+1} \anda M_{N+1}$}
 \label{trees}
\end{figure}

\begin{figure}
\centering
\includegraphics[width=4in,height=2in]{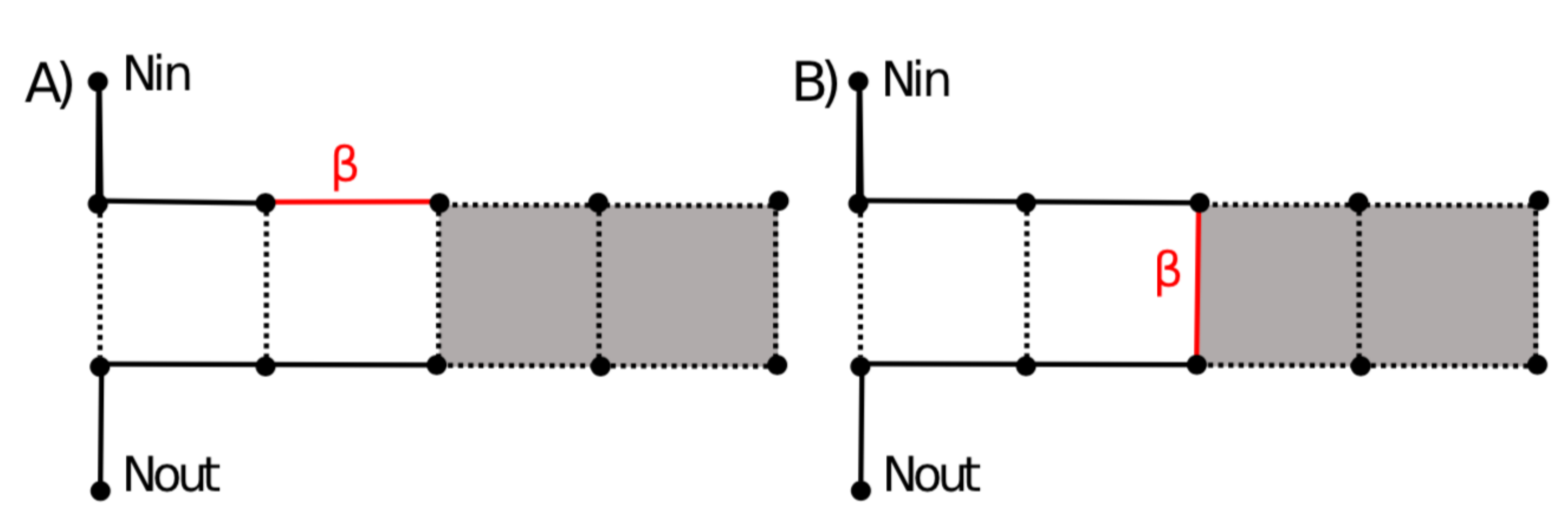}
\caption{${\cal P}_N(\beta)$}
 \label{pnbeta}
\end{figure}

\begin{figure}
\centering
\includegraphics[width=5in,height=3in]{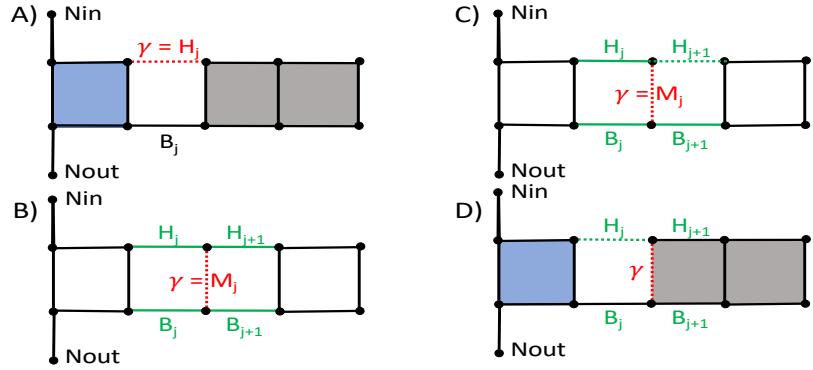}
\caption{$\sigma_N(\gamma)$}
 \label{sigman}
\end{figure}

  \begin{figure}
\centering
\includegraphics[width=5in,height=3in]{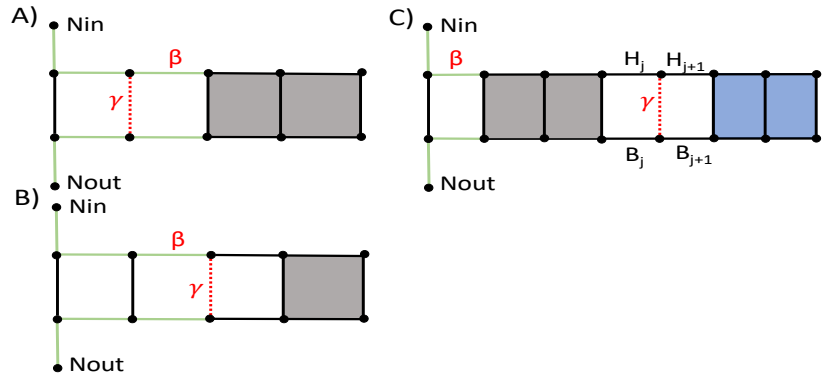}
\caption{${\cal P_N}(\gamma, \beta), \, \beta = H_k, \gamma=M_j$}
 \label{F5}
\end{figure}

\begin{figure}
\centering
\includegraphics[width=5in,height=2in]{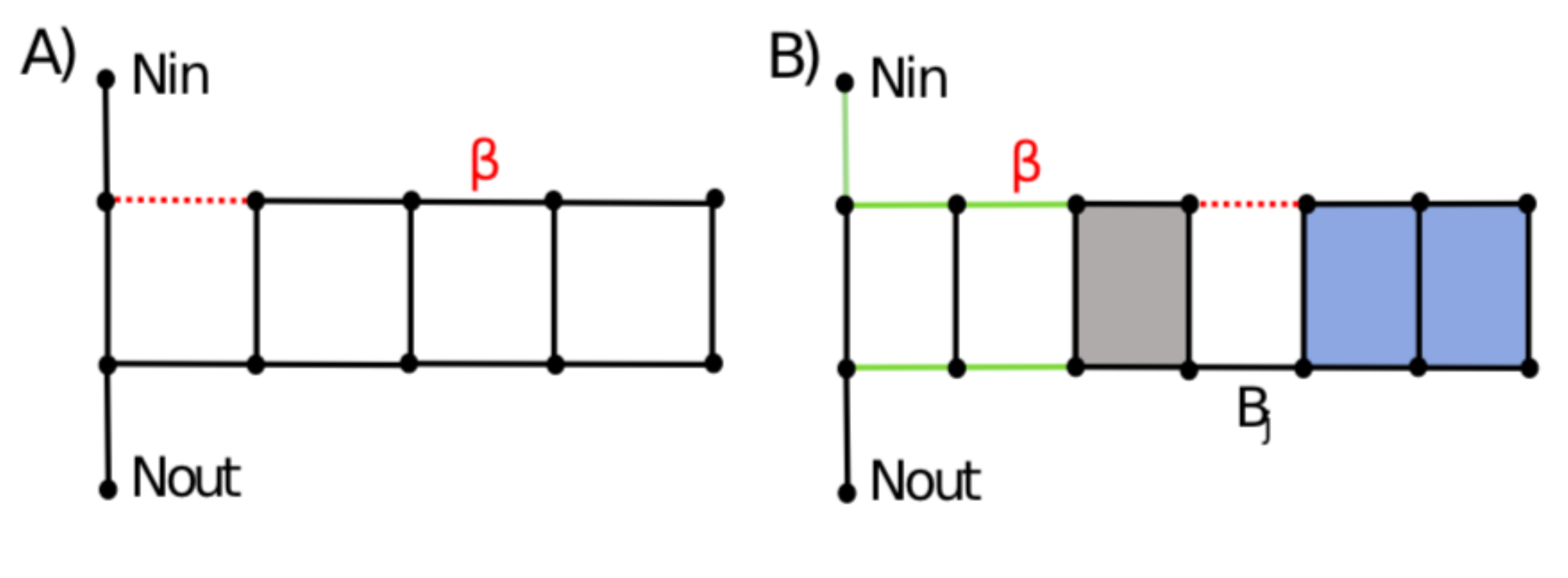}
\caption{${\cal P_N}(\gamma, \beta), \, \beta = H_k, \gamma=B_j$}
 \label{F6}
\end{figure}

\newpage
  
  \bibliographystyle{plain}
  
\bibliography{arxref}

 \end{document}